\documentclass[11pt,a4paper]{article} 
\pdfoutput=1
\usepackage{jheppub}
\usepackage{soul}

\usepackage{amsmath}
\usepackage{bm}
\usepackage{multirow}
\allowdisplaybreaks[1]

\def\l{\left} 
\def\r{\right} 
\def\nl{\nonumber\\} 
\newcommand{\GeV}{\mathinner{\mathrm{GeV}}}

\newcommand\fverb{\setbox\fverbbox=\hbox\bgroup\verb}
\newcommand\fverbdo{\egroup\medskip\noindent%
			\fbox{\unhbox\fverbbox}\ }
\newcommand\fverbit{\egroup\item[\fbox{\unhbox\fverbbox}]}
\newbox\fverbbox

\title{Higgs Portal Vector Dark Matter : Revisited}

\author[a]{Seungwon Baek,} 
\author[a]{P. Ko,}
\author[a]{Wan-Il Park}
\author[a]{and Eibun Senaha}

\affiliation[a]{School of Physics, KIAS, \\ Seoul 130-722, Korea}

\emailAdd{sbaek1560@gmail.com}
\emailAdd{pko@kias.re.kr}
\emailAdd{wipark@kias.re.kr}
\emailAdd{senaha@kias.re.kr}

\arxivnumber{}

\abstract{
We revisit the Higgs portal vector dark matter model including a hidden sector
Higgs field that generates the mass of the vector dark matter. 
The model becomes renormalizable and has two scalar bosons, the mixtures 
of the standard model (SM) Higgs and the hidden sector Higgs bosons. 
The strong bound from direct detection such as XENON100 is evaded due to the 
cancellation mechanism between the contributions from two scalar bosons.  
As a result, the model becomes  still viable in large range of dark matter mass,  
contrary to some claims in the literature.  The Higgs properties are also affected, 
the signal strengths for the Higgs boson search being universally suppressed 
relative to the SM value, which could be tested at the LHC in the future.
}
\begin{document} 
\maketitle
\section{Introduction\label{sec:intro}}

The so-called Higgs portal cold dark matter (CDM) model is an interesting 
possibility for the nonbaryonic dark matter of the universe.  The dark matter
fields are assumed to be the standard model (SM) gauge singlets, and could be a scalar
($S$), a singlet fermion ($\psi$) or a vector boson ($X$) depending on their spin.
The Lagrangian of these CDM's are usually taken as \cite{EFT,EFT2,EFT_DLMQ,EFT_pseudo}
\begin{eqnarray}
{\cal L}_{\rm scalar} & = & \frac{1}{2} \partial_\mu S \partial^\mu S 
- \frac{1}{2} m_S^2 S^2 - \frac{\lambda_{HS}}{2}H^\dagger H S^2 
- \frac{\lambda_S}{4 \!} S^4
\label{EFT:SDM}
\\
{\cal L}_{\rm fermion} & = & \overline{\psi} \left[ 
i \gamma \cdot \partial - m_\psi \right] \psi 
-  \frac{\lambda_{H\psi}}{\Lambda} H^\dagger H ~\overline{\psi} \psi
\label{EFT:SFDM}
\\
{\cal L}_{\rm vector} & = & - \frac{1}{4} X_{\mu\nu} X^{\mu\nu} 
+ \frac{1}{2} m_X^2 X_\mu X^\mu 
+ \frac{1}{4} \lambda_X (X_\mu X^\mu )^2 
+ \frac{1}{2} \lambda_{HX} H^\dagger H {X_\mu X^\mu} .
\label{EFT:VDM}
\end{eqnarray}
Dark matter fields $(S,\psi, X_\mu)$ are assumed to be odd under some discrete 
$Z_2$ symmetry: $(S,\psi, X_\mu) \rightarrow - (S,\psi, X_\mu)$  in order to guarantee 
the stability of CDM.  This symmetry removes the kinetic mixing between the 
$X_{\mu\nu}$  and the $U(1)_Y$ gauge field $B^{\mu\nu}$, making $X_\mu$ stable. 

The scalar CDM model (\ref{EFT:SDM}) is satisfactory both theoretically 
and phenomenologically, as long as $Z_2$ symmetry is unbroken.
The model is renormalizable and can be considered to high energy scale as long
as the Landau pole is not hit.  A large region of parameter space is still
allowed by the relic density and direct detection experiments~\cite{EFT_DLMQ}.
On the other hand, the other two cases have problems. 

Let us first consider the fermionic CDM model (\ref{EFT:SFDM}). 
This model is nonrenormalizable, and has to be UV completed.  The simplest way to achieve the UV completion of (\ref{EFT:SFDM})  is to introduce a real singlet scalar field as proposed in 
Ref.~\cite{Kim:2008pp,SFDM1} by some of us. 
We observed that there are two Higgs-like scalar bosons which interfere
destructively in the spin-independent cross section of the singlet fermion CDM 
on nucleon.   The strong constraint from direct detection experiments such as 
XENON100 \cite{XENON100_2012} or CDMS \cite{Ahmed:2011gh} can be relaxed significantly.
On the other hand, the effective field theory (EFT)  based on the Lagrangian 
(\ref{EFT:SFDM}) is strongly constrained for DM masses below about 2 TeV 
\cite{EFT,EFT2,EFT_DLMQ},  although the EFT with pseudo-scalar Higgs portal 
suggested in \cite{EFT_pseudo} can be still consistent with the current direct search 
bound even for light DM masses. 
The decoupling of the 2nd scalar boson occurs rather slowly, since the mass mixing 
between the SM Higgs boson and the new singlet scalar is due to the dim-2 operator 
\cite{SFDM1}.   
Also the mixing between two scalar bosons makes the signal strength of two physical
Higgs-like bosons less than one, and make it difficult to detect both of them at the LHC.
Since there is now an evidence for a new boson at 125 GeV at the LHC \cite{:2012gk,:2012gu}, the 2nd scalar boson in the singlet fermion DM model is very difficult to be 
observed at the LHC because  its signal strength is  much less than 1
~\cite{SFDM1,SFDM2}.   
Also an extra singlet scalar solves the vacuum instability problem for 
$m_H = 125$ GeV in the SM~\cite{Lebedev:2012zw,EliasMiro:2012ay,SFDM2}, 
making the electroweak (EW) vacuum stable up to Planck scale for 
$m_t = 173.2$ GeV.
These phenomena would be very generic in general hidden sector DM models 
\cite{generic}.  In short, it is very important to consider a renormalizable 
model when one considers the phenomenology of a singlet fermion CDM. 

Now let us turn to the Higgs portal vector dark matter described by (\ref{EFT:VDM}) 
\cite{EFT,EFT2,EFT_DLMQ}. 
This model is very simple, compact and seemingly renormalizable since it has only
dim-2 and dim-4 operators.  However, it is not really renormalizable and  violates unitarity, 
just like the intermediate vector boson model for massive weak gauge bosons before 
Higgs mechanism was developed. 
The Higgs portal VDM model based on (\ref{EFT:VDM})  is a sort of an effective 
Lagrangian which has to be UV completed. 
It lacks  the dark Higgs field, $\Phi (x)$, that would generate the dark gauge field mass
and will mix with the SM Higgs field, $H(x)$, after $U(1)_X$ symmetry breaking. 
Therefore the model (\ref{EFT:VDM}) does not capture dark matter or Higgs boson 
phenomenology correctly.   It is the purpose of this work to propose a simple UV 
completion of the model (\ref{EFT:VDM})  with hidden sector $U(1)_X$ gauge 
symmetry (see also Ref.~\cite{Farzan:2012hh} for a similar approach), and deduce 
the correct phenomenology of vector CDM and two Higgs-like scalar bosons. 
Vector dark matter models in extended gauge symmetries can be found 
in~\cite{VDM}.  Qualitative aspects of our model are similar
to those presented in Ref.s~\cite{SFDM1,SFDM2}, although there are some 
quantitative differences due to the vector nature of the CDM. 

This work is organized as follows. In Sec.~2, we define the model by including the 
hidden sector Higgs field that generates the vector dark matter mass by the usual 
Higgs mechanism.   
Then we present dark matter and collider phenomenology in the 
following section. We also compare the full theory with the EFT, and discuss
the region in which the EFT approach is valid.
The vacuum structure and the vacuum stability issues are discussed
in Sec.~4, and the results are summarized in Sec.~5.

\section{The model Lagrangian for vector dark matter}


Let us consider a vector boson dark matter, $X_\mu$, which is assumed to be
a gauge boson associated with Abelian dark gauge symmetry $U(1)_X$. 
The simplest model will be without any matter fields charged under $U(1)_X$ 
except for a complex scalar, $\Phi$, whose VEV will generate the mass for 
$X_\mu$ (see also Ref.~\cite{Farzan:2012hh}):
\begin{eqnarray}
{\cal L}_{VDM} & = & - \frac{1}{4} X_{\mu\nu} X^{\mu\nu} +  
(D_\mu \Phi)^\dagger (D^\mu \Phi) 
- \lambda_\Phi \l( \Phi^\dagger  \Phi - \frac{v_\Phi^2}{2} \r)^2
\nonumber \\
& & - \lambda_{H\Phi} \l(H^\dagger H - \frac{v_H^2}{2}\r) 
\l(\Phi^\dagger \Phi - \frac{v_\Phi^2}{2}\r) \ ,
\label{eq:full_theory}
\end{eqnarray}
in addition to the SM Lagrangian which includes the Higgs potential term
\begin{equation}
\Delta {\cal L}_{\rm SM} = 
- \lambda_H \l( H^\dagger  H - \frac{v_H^2}{2} \r)^2.
\end{equation}
The covariant derivative is defined
as 
\[
D_\mu \Phi = (\partial_\mu + i g_X Q_\Phi X_\mu) \Phi ,
\]
where $Q_\Phi \equiv Q_X(\Phi)$ is the $U(1)_X$ charge of $\Phi$ and we will take $Q_\Phi=1$ throughout
the paper.

Assuming that the $U(1)_X$-charged complex scalar $\Phi$ develops a nonzero VEV, 
$v_\Phi$, and thus breaks $U(1)_X$ spontaneously, 
\[
 \Phi  = \frac{1}{\sqrt{2}} \left( v_\Phi+ \varphi(x) \right) .
\]
Therefore the Abelian vector boson $X_\mu$ gets mass $M_X = g_X |Q_\Phi| v_\Phi$, 
and the hidden sector Higgs field (or dark Higgs field) $\varphi (x)$ will mix with 
the SM Higgs field $h(x)$ through Higgs portal of the $\lambda_{H\Phi}$ term.
The mixing matrix $O$ between the two scalar fields is defined as
\begin{equation}
\left(
 \begin{array}{c}
  h \\ \varphi
 \end{array}
\right)
= O 
\left(
 \begin{array}{c}
  H_1 \\ H_2
 \end{array}
\right)
\equiv  
\left(
 \begin{array}{cc}
  c_\alpha & s_\alpha \\
  -s_\alpha & c_\alpha 
 \end{array}
\right)
\left(
 \begin{array}{c}
  H_1 \\ H_2
 \end{array}
\right),
\end{equation}
where $s_\alpha (c_\alpha) \equiv \sin\alpha (\cos\alpha)$, $h, \varphi$ 
are the interaction eigenstates and $H_i (i=1,2)$ are the
mass eigenstates with masses $m_i$. The mass matrix in the basis $(h,\varphi)$ can be written 
in terms either of Lagrangian parameters or of the physical parameters as follows:
\begin{equation}
\left(
\begin{array}{cc}
2 \lambda_H v_H^2 & \lambda_{H\Phi} v_H v_\Phi \\ 
\lambda_{H\Phi} v_H v_\Phi & 2 \lambda_\Phi v_\Phi^2
\end{array}
\right) =
\left(
\begin{array}{cc}
m_1^2 c_\alpha^2 + m_2^2 s_\alpha^2 & (m_2^2 -m_1^2 )s_\alpha c_\alpha \\ 
(m_2^2 -m_1^2 )s_\alpha c_\alpha & m_1^2 s_\alpha^2 + m_2^2 c_\alpha^2
\end{array}
\right).
\label{eq:mass_matrix}
\end{equation}

\section{Phenomenology}

\subsection{Dark matter phenomenology}
The observed present cold dark matter density,
$\Omega_{\rm CDM} h^2 \simeq 0.1123 \pm 0.0035$~\cite{WMAP}, is approximately 
related to the thermally averaged annihilation cross section at freeze-out temperature, 
$\langle \sigma v \rangle_{\rm fz}$, as
\begin{equation}
\Omega_{\rm CDM} h^2 = \frac{3 \times 10^{-27} {\rm cm^3/s }}
{\langle \sigma v \rangle_{\rm fz}}.
\end{equation}
So we require $\langle \sigma v \rangle_{\rm fz} \approx 3 \times 10^{-26} {\rm cm^3/s}$
to obtain the correct relic density.  We have used the micrOmegas v.2.4.5 \cite{micromegas} 
to calculate thermal relic density and direct detection cross section of the VDM in our model.

\begin{figure}
\centering
\includegraphics[width=0.7\textwidth]{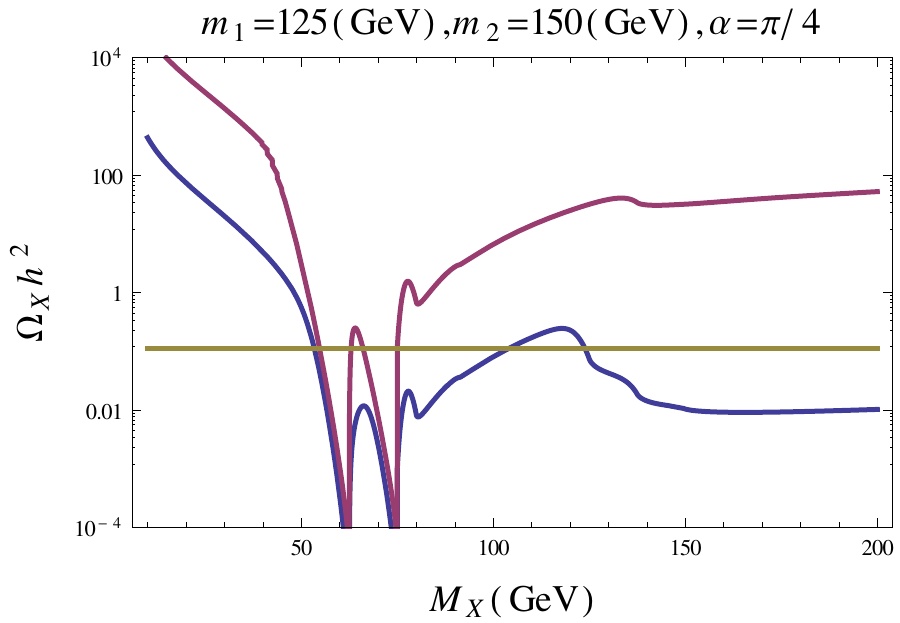}
\caption{The thermal relic density $\Omega_X h^2$ of the vector dark matter
as of function of the dark matter mass, $M_X$. For this plot we fixed
$m_1 = 125~\GeV, m_2 = 150~\GeV, \alpha=\pi/4$ and the purple (blue) line
corresponds to $g_X =0.05 ~(0.5)$. The horizontal line is the central value
of the current relic density $\Omega_X h^2 = 0.1123$~\cite{WMAP}.
}
\label{fig:Om_MDM}
\end{figure}

In Fig.~\ref{fig:Om_MDM} we show the thermal relic density as a function of the 
dark matter mass, $M_X$.
For this plot we fixed
$m_1 = 125~\GeV, m_2 = 150~\GeV, \alpha=\pi/4$ and the purple (blue) line
corresponds to $g_X =0.05 ~(0.5)$.
We can see two resonance dips at $M_X=m_i/2$ ($i=1,2$). 
The VDMs can annihilate into the SM particles in the S-wave state,
which is different from the singlet fermionic dark matter case studied
in~\cite{SFDM1} where the annihilation occurs in the P-wave state.
As a result the annihilation cross section for the vector dark matter is generally 
${\cal O}(10 - 100)$ larger than that of the SFDM. 
And the current relic density can be explained more easily even at non-resonance region. 
(See the blue line in Fig.~\ref{fig:Om_MDM}.)
The difference between the two curves becomes larger for $M_X >125$~GeV.
This is because the channels $X X \to H_i H_j$ ($i,j=1,2$) which begin to
open for $M_X >125$~GeV are sensitive to $g_X$ and they give larger annihilation
cross sections as the coupling $g_X$ increases.

\begin{figure}
\centering
\includegraphics[width=0.7\textwidth]{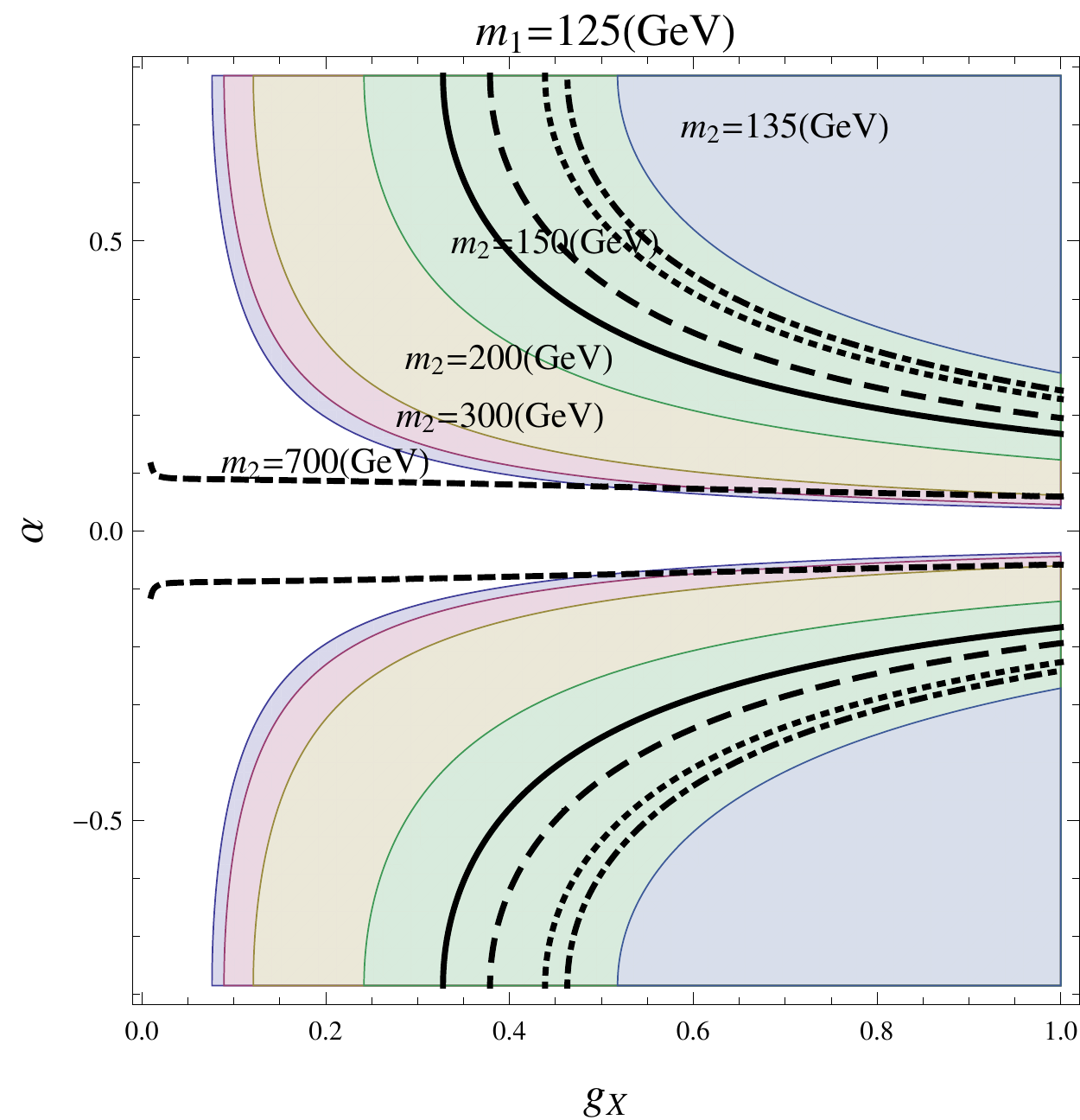}
\caption{The excluded region in the $(g_X,\alpha)$-plane.
Each colored region is excluded by XENON100 direct detection experiment for the
$m_2$ value given in the plot. We fixed $M_X = 70 \GeV, m_1=125 \GeV$. The black 
solid (dashed, long-dashed, dotted, dot-dashed) curve 
corresponds to $\Omega_X h^2=0.1123$ for $m_2=135 (150, 200, 300, 700)$~GeV. 
Therefore, the VDM as light as $M_X = 70$ GeV is 
allowed by both the relic density and the XENON100 
constraints either by the cancellation mechanism for $m_2 = 135$ GeV
or by the resonant annihilation for $m_2=150$ GeV.  The entire region is also allowed by the 
$S, T, U$-parameters at 99\% confidence level except that only the range $(-0.63,0.63)$
and $(-0.42,0.42)$ of $\alpha$ are allowed for $m_2=300$ GeV and $m_2=700$ GeV, 
respectively~\cite{SFDM1}. 
}
\label{fig:gx_alpha}
\end{figure}

One important effect when considering the full theory, 
which we found in Ref.~\cite{SFDM1},  is that a generic cancellation occurs 
in the dark matter and nucleon scattering amplitude, which can not be observed
in the effective Lagrangian approach.\footnote{In general the cancellation 
mechanism can also work in the annihilation 
process  for the relic density. However, the different decay widths for the $H_1$ 
and $H_2$ and/or  other processes such as annihilations into scalar particle pairs 
makes it less effective  than in the direct detection process.
As a result, the annihilation process and the direct detection process are not 
strictly proportional to each other in our scenario.}
This is because the transformation matrix between the interaction eigenstates 
and the mass  eigenstates in the scalar sector is an orthogonal matrix.
The dark matter and nucleon elastic scattering cross section is proportional to
the following factor:
\begin{equation}
 \sigma_p \propto  \l|\sum_{i=1,2} \frac{O_{hi} O_{\varphi i}}{q^2-m_i^2}\r|^2,
\end{equation}
where $q$ is the momentum transfer of the dark matter.
When $m_1 \approx m_2$ or $|q^2| \gg m_i^2$, we have $\sigma_p \approx 0$ due to the orthogonality
of the mixing matrix $O$. This cancellation phenomenon is quite similar to the GIM-mechanism \cite{Glashow:1970gm}
in the quark (or lepton) flavor violating neutral current processes. 
In Fig.~\ref{fig:gx_alpha}, we show the excluded region
in the $(g_X,\alpha)$-plane by the non-observation of dark matter by the 
XENON100 which currently gives the strongest bound on the dark matter direct detection cross
section~\cite{XENON100_2012}. 
Each colored region is excluded by XENON100 direct detection experiment for the
$m_2$ value given in the plot. We fixed $M_X = 70 \GeV, m_1=125 \GeV$ for the plot. 
The black solid (dashed, long-dashed, dotted, dot-dashed) curve corresponds to 
$\Omega_X h^2=0.1123$ for  $m_2=135~ (150, 200, 300, 700)$~GeV. 
The case $m_2 = 150$ GeV is close to the resonance ($m_2 = 2M_X$)
and shows quite different behavior from the other cases.
So the VDM as light as $M_X = 70$ GeV, even if it is off the resonance region, 
can be consistent 
with both the relic density 
and the XENON100 experiment  by the cancellation mechanism 
when $H_2$ is light. This can be compared with the EFT approach based on the
Lagrangian (\ref{EFT:VDM}) where $M_X \lesssim 300$ GeV is already excluded by the 
direct search limit~\cite{EFT2} (See also the blue line in Fig.~\ref{fig:sigp} (a)).
The entire region is also allowed by the 
electroweak precision $S, T, U$-parameters at 99\% confidence level
except that only the range $(-0.63,0.63)$
and $(-0.42,0.42)$ of $\alpha$ are allowed for $m_2=300$ GeV and $m_2=700$ GeV, 
respectively~\cite{SFDM1}.

\begin{figure}
\centering
\includegraphics[width=0.7\textwidth]{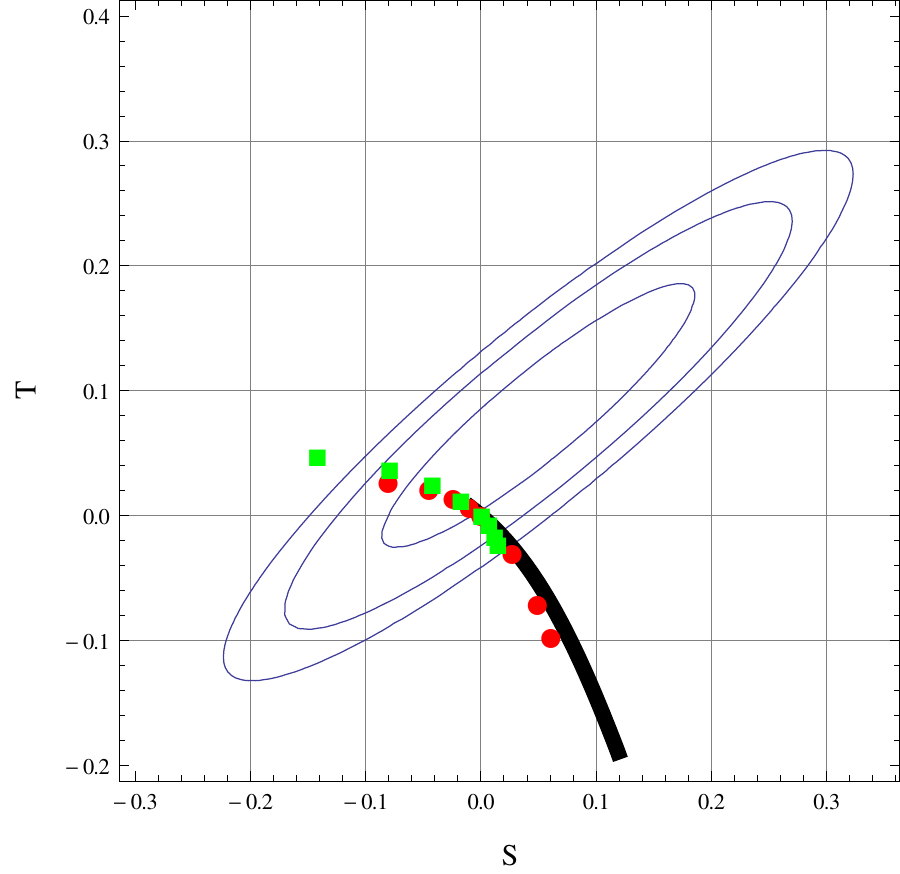}
\caption{The predictions of $(S,T)$-parameters in our model for $(m_1,m_2)=
(25,125)$, $(50,125)$, $(75,125)$, $(100,125)$, $(125,125)$, $(125,250)$,
 $(125,500)$, $(125,750)$ GeV from above.
The green (red) dots are for $\alpha=45^\circ (20^\circ)$.
The thick black line is the prediction of the SM with the $m_H$ in the range
$[125,720]$ GeV. The ellipses represent 68, 95, 99\% CL experimental lines from inside out.
}
\label{fig:ST}
\end{figure}

The predictions of our model on the $S,T$ parameters assuming $U=0$
are shown in Fig.~\ref{fig:ST} for the choices  
$(m_1,m_2)=(25,125)$, $(50,125)$, $(75,125)$, $(100,125)$, $(125,125)$, $(125,250)$,
 $(125,500)$, $(125,750)$ GeV  from above.
The green (red) dots are for $\alpha=45^\circ (20^\circ)$.
The thick black line is the prediction of the SM with the $m_H$ in the range
$[125,720]$ GeV. The ellipses represent 68, 95, 99\% CL experimental lines from inside out.

%

\subsection{Collider phenomenology}
Since the scalar sector is extended, the Higgs phenomenology is different
from that of the SM. In this subsection we study the possibility that the second
Higgs which our model predicts could be discovered at the LHC. We will also
see that the combination of the collider signatures and the DM direct searches
is robust enough to exclude or confirm our model in the on-going LHC and the 
next generation DM direct detection experiments.

The signal strength of a scalar boson $H_{i=1,2}$ defined as
\begin{equation}
 r_i \equiv \frac{\sigma(p p \to H_i) B(H_i \to f_{\rm SM})}{[\sigma(p p \to H_i) 
 B(H_i \to f_{\rm SM})]^{\rm SM}}
\end{equation}
can be measured at the LHC. Here $i=1,2$ and $f_{\rm SM}$ is a specific SM final state 
which the scalar boson $H_i$ can decay into. In our model it can be written in terms 
of  $\Gamma_i^{\rm tot,SM}$ ($i=1,2$) which is the total decay width of $H_i$ 
in the SM assuming $H_i$ is a pure SM Higgs
and $\Gamma_i^{\rm tot}$ which is the total decay  width of $H_i$ in our model 
\cite{SFDM1,SFDM2}:
\begin{equation}
r_i = O_{hi}^4 \frac{\Gamma_i^{\rm tot,SM}}{\Gamma_i^{\rm tot}},
\label{eq:r_i}
\end{equation}
where $O_{h1}=c_\alpha, O_{h2} = s_\alpha$.
The total decay widths can be decomposed as
\begin{eqnarray}
\Gamma_1^{\rm tot} &=& c_\alpha^2 \Gamma_1^{\rm tot, SM} +s_\alpha^2 \Gamma_1^{\rm tot, hid}, \nl
\Gamma_2^{\rm tot} &=& s_\alpha^2 \Gamma_2^{\rm tot, SM} +c_\alpha^2 \Gamma_2^{\rm tot, hid}
  +\Gamma(H_2 \to H_1 H_1), 
\label{eq:Gamma_i}
\end{eqnarray}
where $\Gamma_i^{\rm tot,hid}$ is the total decay width of $H_i$ into the hidden
sector assuming $H_i$ is a pure SM-singlet scalar. The channel
$H_2 \to H_1 H_1$ opens when $m_2 > 2 m_1$.
From the eqs.~(\ref{eq:r_i}) and (\ref{eq:Gamma_i}) it is obvious that $r_i <1$ 
in our model. 
Therefore if the excess of the signal strength in some channels 
like $H \to \gamma\gamma$ above
the SM prediction at the LHC remains in the future data, our model will
either be excluded or need to be extended (two Higgs doublet portal to a hidden
sector dark matter, for example).
From $r_1+r_2 <1$~\cite{SFDM1,SFDM2} we  obtain  $r_2 < 0.3$ for the second 
Higgs boson, when we identify the observed new boson at 125 GeV (whose  
 signal strength is greater than 0.7 at 2$\sigma$ level~\cite{ATLAS_Higgs}
\footnote{We used only the ATLAS value since there is no combined result.
The corresponding value for the CMS can be found in~\cite{CMS_Higgs}}) 
as one of the two Higgs-like scalar bosons in our model.

\begin{figure}
\centering
\includegraphics[width=0.7\textwidth]{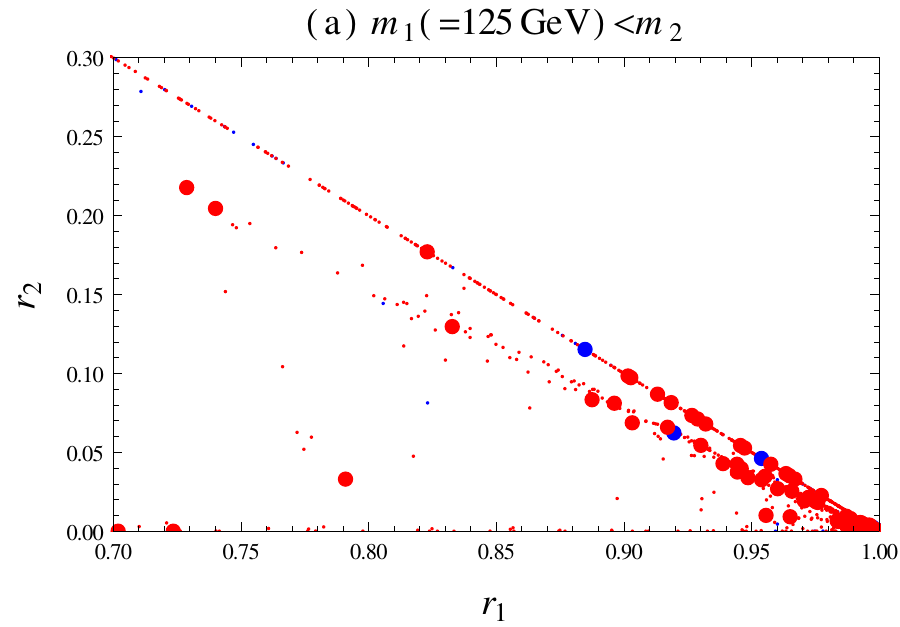}
\includegraphics[width=0.7\textwidth]{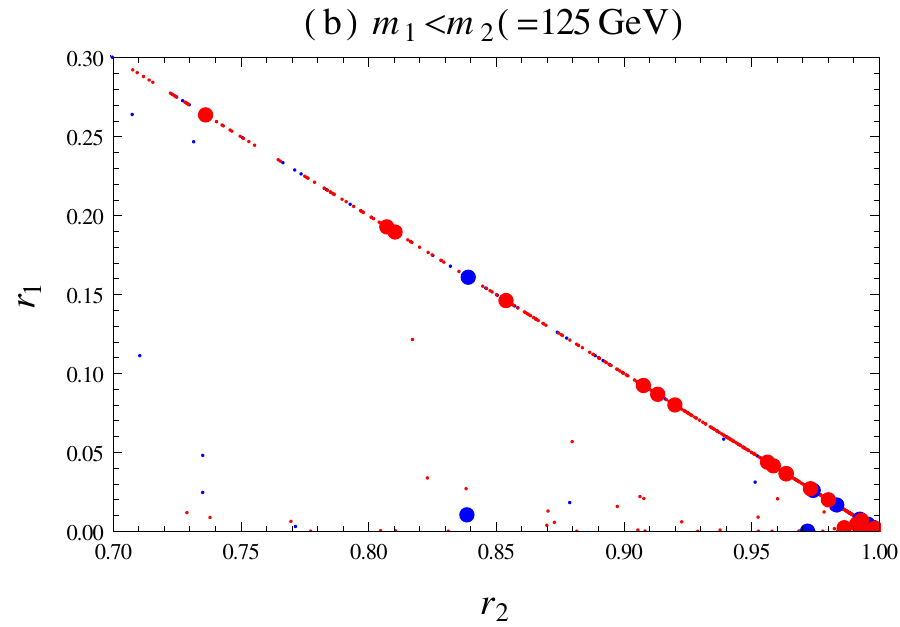}
\caption{The scatter plot in (a) $(r_1,r_2)$ for 
$m_1 (= 125 ~{\rm GeV)} < m_2$ and (b) $(r_2,r_1)$ for 
$m_1 < m_2 (= 125~ {\rm GeV}) $.
The big (small) points (do not) satisfy the WMAP relic density constraint 
within 3 $\sigma$, while the red-(blue-)colored points can (cannot) be
probed at the planned XENON1T direct detection experiment.
}
\label{fig:r1r2}
\end{figure}

The correlation between $r_1$ and $r_2$ can be seen in Fig.~\ref{fig:r1r2} where
we show only the region $r_1>0.7$.
For this plot we scanned the parameters $g_X$, $M_X$, $\alpha$, $m_2$ in the 
range, $0 <g_X <1$, $10~{\rm GeV} <M_X< 1000 {\rm GeV}$, 
$-\pi/2 <\alpha < \pi/2$,  $ m_1(=125 \GeV) <m_2 <2000 \GeV$ for the panel (a),  
and $ 10 \GeV <m_1 <m_2 (=125 \GeV)$ for the panel (b). 
All the points pass the constraints:
$\Omega_X h^2 < 0.1228$ (the 3$\sigma$ upper bound of the relic density), 
the upper bound on the XENON100 direct detection cross section,
and the bound on the $S,T$-parameters at 99\% CL.
The big (small) points (do not) satisfy the WMAP relic density constraint 
within 3 $\sigma$, while the red-(blue-)colored points can (cannot) be
probed at the planned XENON1T direct detection experiment~\cite{XENON1T}.
In both plots, the big red points on the straight line, $r_1 + r_2 =1$, are those with 
$H_i \to X X$ and $H_2 \to H_1 H_1$ suppressed.
In the panel (a), the sizable contribution from the $H_2 \to H_1 H_1$ channel 
allows the big red points below the $r_1 + r_2 =1$ line.

\begin{figure}
\centering
\includegraphics[width=0.7\textwidth]{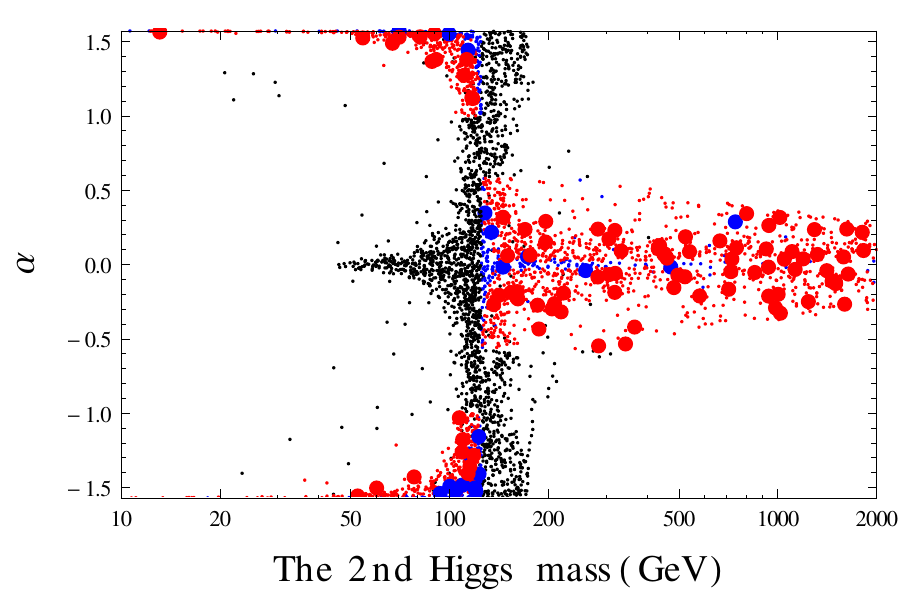}
\caption{The allowed mixing angle $\alpha$ as a function of the second Higgs 
mass.   We fixed the SM-like Higgs mass to be 125 GeV.
The big (small) points (do not) satisfy the WMAP relic density constraint 
within 3$\sigma$, while the red-(blue-)colored points can (cannot) be probed
at the planned XENON1T experiment. The black points are excluded by the LHC 
Higgs search, {\it i.e.} $r < 0.7$.
}
\label{fig:m2_alpha}
\end{figure}

In Fig.~\ref{fig:m2_alpha}, we show  the allowed mixing angle $\alpha$ as a function 
of the second Higgs mass.   We fixed the SM-like Higgs mass to be 125 GeV.
Color scheme is the same as Fig.~\ref{fig:r1r2} except that black points are excluded 
by the LHC Higgs search, {\it i.e.} $r < 0.7$.
We can see the maximal mixing angle $\alpha=\pi/4$ (black points near 
$m_2 \approx 125 \GeV$) is excluded by the LHC Higgs search.
Also the light scalar with mass less than 125 GeV, if exists, should be
singlet-like.

\begin{figure}
\centering
\includegraphics[width=0.8\textwidth]{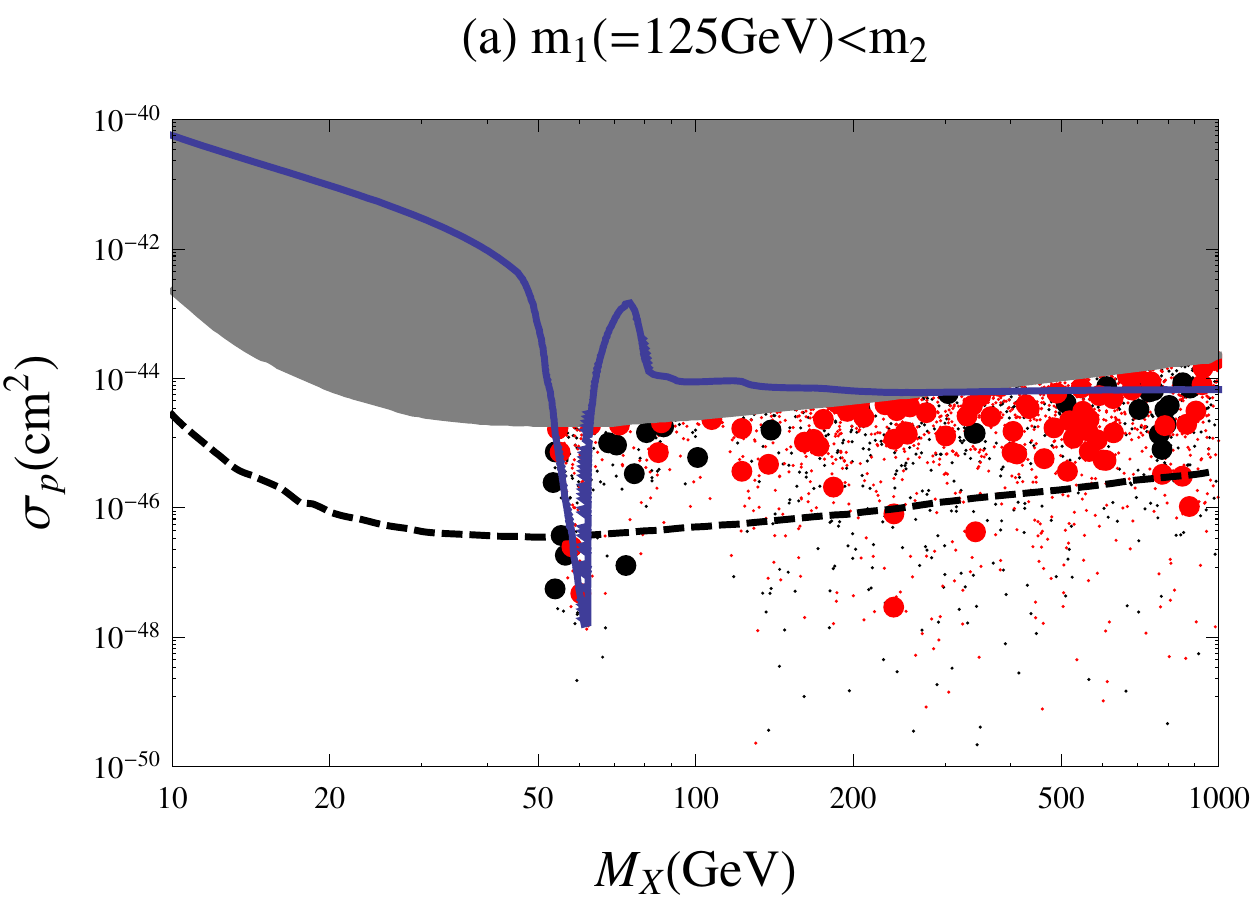}
\includegraphics[width=0.8\textwidth]{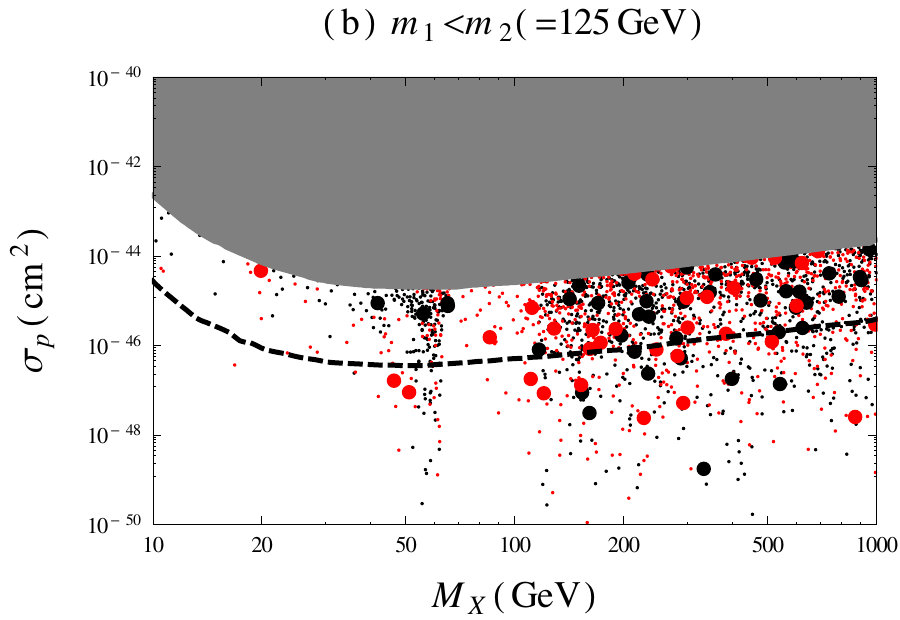}
\caption{The scatter plot of $\sigma_p$ as a function of $M_X$.
The big (small) points (do not) satisfy the WMAP relic density constraint 
within 3 $\sigma$, while the red-(black-)colored points gives
$r_1 > 0.7 (r_1 <0.7)$. The gray region is excluded by the XENON100 experiment.
The dashed line denotes the sensitivity of the next XENON experiment, XENON1T.
The solid blue line in panel (a) represents the prediction of the EFT
approach in (\ref{EFT:VDM}).
}
\label{fig:sigp}
\end{figure}

In Fig.~\ref{fig:sigp}, we show a scatter plot of $\sigma_p$ as a function of $M_X$.
The big (small) points (do not) satisfy the WMAP relic density constraint 
within 3 $\sigma$, while the red-(black-)colored points gives
$r_1 > 0.7 (r_1 <0.7)$. The Gray region is excluded by the XENON100 experiment.
The dashed line denotes the sensitivity of the next XENON experiment, XENON1T.
We note that many points are still allowed by the WMAP relic density
constraint, the XENON100 direct detection experiment, and also by the constraint 
$r_1>0.7$ which is in the ball park of the LHC Higgs search bound. On the other hand, 
the effective field theory approach considered in Ref.s~\cite{EFT}  strongly constrains 
the vector dark matter scenario. 
We can also see that there is no point below about $M_X \approx 50$ GeV in Fig.~\ref{fig:sigp} (a).
It is because the Higgs exchanged dark matter annihilation channel
does not allow the resonance and the relic density is larger than the WMAP measurement.
Most of the big red points are within the reach of the XENON1T sensitivity, 
and our model can be tested in the next generation dark matter detection experiment.

\subsection{The EFT as a limit of the full theory for $m_2 \to \infty$}
In this subsection we consider the EFT in (\ref{EFT:VDM}) as a limit of
the full theory in (\ref{eq:full_theory}) when $m_2 \to \infty$. We keep finite
the full theory parameters: $\lambda_H$, $\lambda_\Phi$, $\lambda_{H\Phi}$ and
$M_X(=g_X v_\Phi)$, while taking $v_\Phi \to \infty$.
We trade $\lambda_H$ for the experimentally measured $m_1$ using the relation
\begin{equation}
2 \lambda_H v_H^2 = m_1^2 + \frac{(\lambda_{H\Phi} v_H v_\Phi)^2}{2 \lambda_\Phi v_\Phi^2-m_1^2}.
\end{equation}
For large $v_\Phi$, $m_2$ ($\alpha$) is proportional to
$v_\Phi$ ($1/v_\Phi$).
The light dark matter ($M_X \ll m_2$) is possible when $g_X^2 \ll \lambda_\Phi$.
In other words, we should note that the EFT is valid only when $m_2 \to \infty$,
$\alpha \to 0$, $g_X^2 \ll \lambda_\Phi$ and it is a very restricted region.
The term $H^\dagger H X_\mu X^\mu$ can be generated both at tree- and  
loop-level~\cite{EFT2}.
Considering the tree-level diagram only, we get
\begin{equation}
\lambda_{HX} = -\frac{2 M_X^2 \widetilde{\lambda}_{112}}{m_2^2}, 
\label{eq:eff_coupl}
\end{equation}
where $\widetilde{\lambda}_{112} \equiv \lambda_{112} c_\alpha/v_\Phi$
and $\lambda_{112}$ is the $H_1-H_1-H_2$ coupling constant given by
\begin{equation}
\lambda_{112} = \lambda_{H\Phi} \Big[(c_\alpha^3 -2 c_\alpha s_\alpha^2) v_\Phi 
+(s_\alpha^3 -2 s_\alpha c_\alpha^2) v_H \Big]+6 \lambda_H s_\alpha c_\alpha^2 v_H
+6 \lambda_\Phi c_\alpha s_\alpha^2 v_\Phi.
\label{eq:112}
\end{equation}

The elastic cross section $\sigma_p$ of the VDM $X$ scattering off the proton 
in the full theory is obtained as
\begin{equation}
\sigma_p^{\rm full} = \frac{4 \mu_X^2}{\pi} \left(\frac{g_X s_\alpha c_\alpha m_p}{2 v_H}\right)^2
\left(\frac{1}{m_1^2}-\frac{1}{m_2^2}\right)^2 f_p^2,
\end{equation}
where $\mu_X =M_X m_p/(M_X + m_p)$ and 
$f_p = \sum_{q=u,d,s} f_q^p + 2/9(1-\sum_{q=u,d,s} f_q^p) \approx 0.468$~\cite{Belanger:2008sj}.
The EFT predicts the corresponding cross section to be
\begin{equation}
\sigma_p^{\rm EFT} = 
\frac{4 \mu_X^2}{\pi} \left(\frac{\lambda_{HX} m_p}{4 M_X}\right)^2
\frac{1}{m_h^4} f_p^2.
\end{equation}
Using the relations (\ref{eq:mass_matrix}), (\ref{eq:eff_coupl}), (\ref{eq:112}) 
and identifying $m_1$ 
and $m_h$ with the observed Higgs mass ($\sim$125 GeV) in their respective theories, we
obtain
\begin{equation}
\frac{\sigma_p^{\rm full}}{\sigma_p^{\rm EFT}}=
\left(\frac{\lambda_{H\Phi}}{\widetilde{\lambda}_{112}}\right)^2.
\end{equation}
This ratio approaches to one as $v_\Phi \to \infty$.
\begin{figure}
\centering
\includegraphics[width=0.7\textwidth]{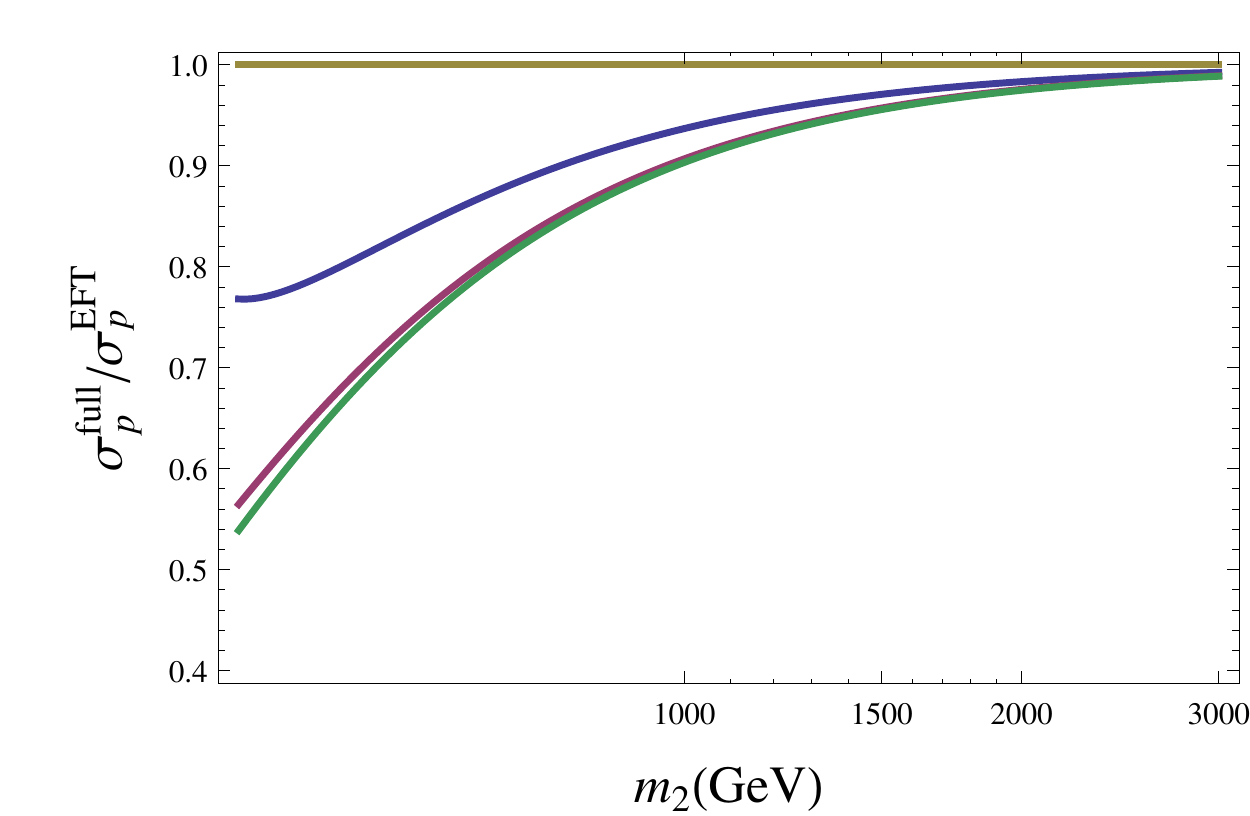}
\caption{The ratio ${\sigma_p^{\rm full}}/{\sigma_p^{\rm EFT}}$ as a function 
of $m_2$ for several values of $\lambda_{H\Phi}$: $\lambda_{H\Phi}=0.5, 0.3, 0.1$ (blue,
purple, green). 
We fix $M_X = 300$ GeV, $m_1 = 125$ GeV $\lambda_\Phi=0.175$.
}
\label{fig:sigp_EFT}
\end{figure}
In Fig.~\ref{fig:sigp_EFT}, we show the ratio ${\sigma_p^{\rm full}}/{\sigma_p^{\rm EFT}}$
as a function of $m_2$ to see how quickly the full theory prediction approaches that of the EFT.
We choose three different values for $\lambda_{H\Phi}$,
$\lambda_{H\Phi}=0.5, 0.3, 0.1$ (blue, purple, green), and fix other parameters:
$M_X = 300$ GeV, $m_1 = 125$ GeV, $\lambda_\Phi=0.175$. We can see that the EFT predictions 
agree well with those of the full theory within a few percent when $m_2 \gtrsim 2000$ GeV.

\begin{figure}
\centering
\includegraphics[width=0.7\textwidth]{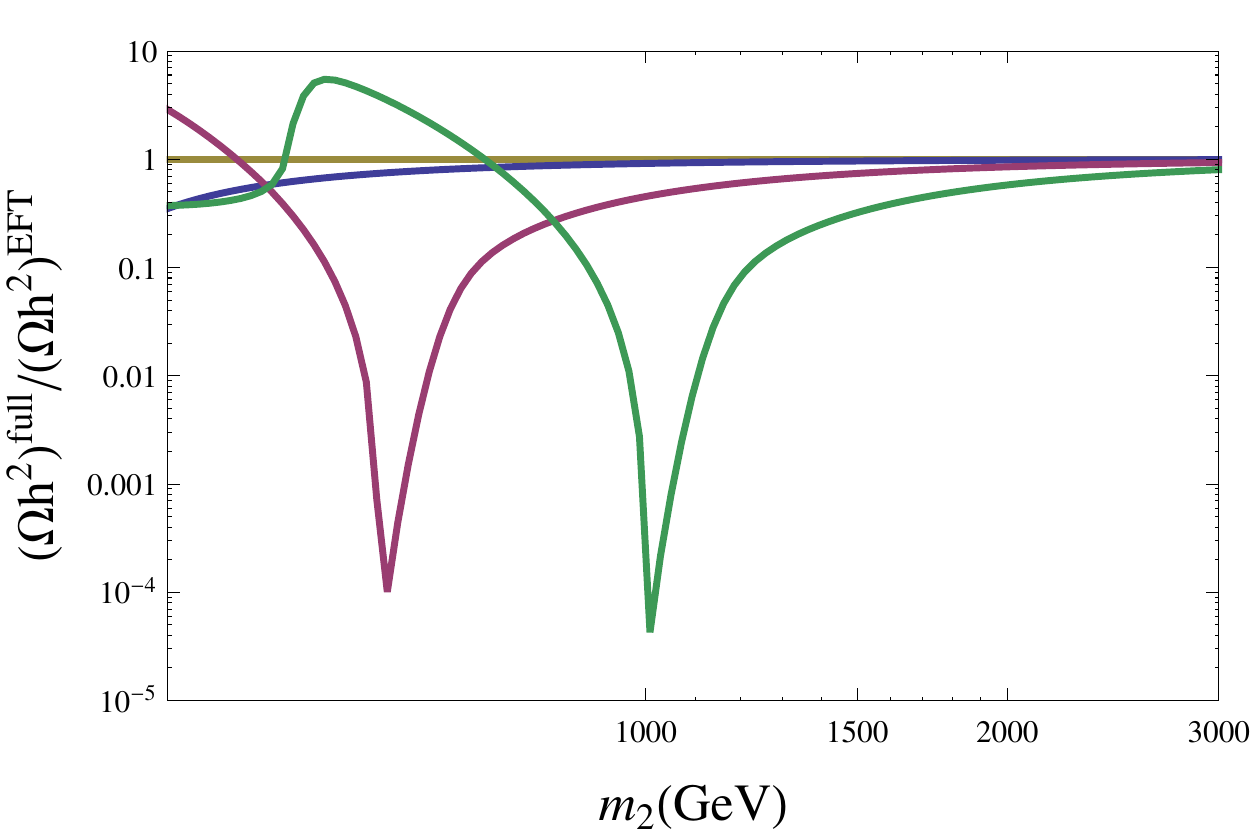}
\caption{The ratio $(\Omega h^2)^{\rm full}/(\Omega h^2)^{\rm EFT}$ as a function 
of $m_2$ for several values of $M_X$: $M_X=150, 300, 500$ GeV (blue,
purple, green). 
We fix $m_1 = 125$ GeV, $\lambda_\Phi=0.175$  and $\lambda_{H\Phi}=0.1$.
}
\label{fig:Om_EFT}
\end{figure}
Fig.~\ref{fig:Om_EFT} shows the ratio of relic density predictions in the
full theory and the EFT, $(\Omega h^2)^{\rm full}/(\Omega h^2)^{\rm EFT}$ , 
as a function of $m_2$. Since the dependence on the coupling 
$\lambda_{H\Phi}$ is not appreciable\footnote{This is partly because the
$X X \to H_2 \to H_1 H_1$ process is important and the amplitudes of which are exactly the same
in both the full theory and the EFT.
}, we take several values of $M_X$ instead:
$M_X =150, 300, 500$ GeV (blue, purple, green), although $M_X=150$ GeV
is already excluded by the direct search limit as can be seen in Fig.~\ref{fig:sigp} (a).
We fix $m_1 = 125$ GeV, $\lambda_\Phi=0.175$ and $\lambda_{H\Phi}=0.1$. 
There is a sharp increase  in the green line at $m_2 \simeq M_X$. 
This is because  the dominant process $X X \to H_2 H_2$ is kinematically closed 
at the point and  the annihilation cross section decreases abruptly in the full theory. 
We can also see the resonance effects of the
full theory. Both effects are absent in the EFT.
We can see that the lighter the DM is, the faster the full theory
approaches the EFT.

\section{Vacuum stability and perturbativity of Higgs quartic couplings}
In this section, we analyze vacuum stability and perturbativity of Higgs quartic couplings.
To make the Higgs potential be bounded-from-below, we require
\begin{align}
\lambda_H>0, \quad \lambda_\Phi>0,\quad -2\sqrt{\lambda_H\lambda_\Phi}<\lambda_{H\Phi},
\label{UB}
\end{align}
where the last condition applies for $\lambda_{H\Phi}<0$. 
We also require 
\begin{align}
\det M_{\rm Higgs}^2 = \det
\left(
	\begin{array}{cc}
	2\lambda_H v_H^2 & \lambda_{H\Phi}v_Hv_\Phi \\
	\lambda_{H\Phi}v_Hv_\Phi & 2\lambda_\Phi v_\Phi^2 
	\end{array}
\right) = (4\lambda_H\lambda_\Phi-\lambda_{H\Phi}^2)v_H^2v_\Phi^2>0.
\label{posi_mHsq}
\end{align}
Since there is additional direction of $\Phi$,  
the Higgs potential can have minima other than our EW vacuum.  
In the following, we investigate whether the EW vacuum is global or not.
We closely follow the analysis done in Ref.~\cite{SFDM2}.

The tree-level effective potential takes the $U(1)_X$ symmetric form
\begin{align}
V_0(\varphi_H, \varphi_\Phi) 
&= \frac{\lambda_H}{4}(\varphi_H^4-2v_H^2\varphi_H^2) 
	+\frac{\lambda_\Phi}{4}(\varphi_\Phi^4-2v_\Phi^2\varphi_\Phi^2)
+\frac{\lambda_{H\Phi}}{4}(\varphi_H^2\varphi_\Phi^2-\varphi_H^2v_\Phi^2-v_H^2\varphi_\Phi^2),
\label{V0}
\end{align}
where $\varphi_H$ and $\varphi_\Phi$ are spacetime-independent classical fields. 
Following the Refs. \cite{Funakubo:2005pu,Cheung:2010ba}, we define 
the various vacua as follows: 
\begin{align}
{\rm EW}&: v_H=246~{\rm GeV},\quad v_\Phi \neq 0, \\
{\rm SYM}&: v_H=v_\Phi=0, \\
{\rm I}&: v_H=0,\quad v_\Phi\neq0, \\
{\rm II}&: v_H\neq0,\quad v_\Phi=0. 
\end{align}

Unlike the general Higgs potential, only nontrivial phase may be the I--phase. 
Such a minimum is given by
\begin{align}
\bar{v}_\Phi =\pm\sqrt{v_\Phi^2+\frac{\lambda_{H\Phi}}{2\lambda_\Phi}v_H^2}.
\label{vPhibar}
\end{align}
The differences of vacuum energies of the I-- and the EW phases is 
\begin{align}
V_0^{\rm (I)}(0, \bar{v}_\Phi)-V_0^{\rm (EW)}(v_H, v_\Phi) 
&=\frac{\lambda_H}{4}v_H^4+\frac{\lambda_{H\Phi}}{4}v_H^2v_\Phi^2
	-\frac{\lambda_\Phi}{4}(\bar{v}_\Phi^4-v_\Phi^4)\nonumber\\
&=\frac{1}{16\lambda_\Phi}(4\lambda_H\lambda_\Phi-\lambda_{H\Phi}^2)v_H^4,
\label{Del_I-EW}
\end{align}
where we have used Eq.~(\ref{vPhibar}) in the second line. 
Therefore, as long as Eqs.~(\ref{UB}) and (\ref{posi_mHsq}) are satisfied, 
the EW vacuum is always the global minimum. 
Note that this is not the case for the generic Higgs potential~\cite{SFDM2}.

Although the EW vacuum is stable at the EW scale, 
its stability up to Planck scale ($M_{\rm Pl}\simeq 1.22\times 10^{19}$ GeV) 
is nontrivial question since a renormalization group (RG) effect 
of the top quark can drive $\lambda_H$ negative at certain high-energy scale, 
leading to an unbounded-from-below Higgs potential or a minimum that may be deeper 
than the EW vacuum.
We will work out this question by solving RG equations with respect to the Higgs quartic couplings
and the $U(1)_X$ gauge coupling.
The one-loop $\beta$ functions of those couplings are listed in Appendix \ref{sec:beta}.
In addition to the vacuum stability, we also take account of the perturbativity of the couplings.
To be specific, we impose $\lambda_i(Q)<4\pi$ ($i=H,H\Phi, \Phi$) 
and $g_X^2(Q)<4\pi$ up to $Q=M_{\rm Pl}$.

\begin{figure}[t]
\centering
\includegraphics[width=10cm]{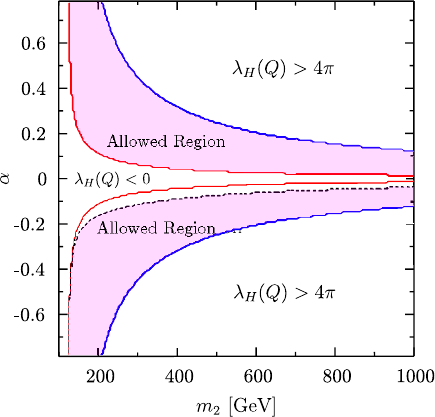}
\caption{
The vacuum stability and perturbativity constraints in the $\alpha$-$m_2$ plane.
We take $m_1=125$ GeV, $g_X^{}=0.05$, $M_X=m_2/2$ and $v_\Phi=M_X/(g_XQ_\Phi)$.
}
\label{fig:RG_al_m2}
\end{figure}
Fig.~\ref{fig:RG_al_m2} shows the vacuum stability and the perturbativity constraints in the $\alpha$-$m_2$ plane.
We take $m_1=125$ GeV, $g_X^{}=0.05$, $M_X=m_2/2$ and $v_\Phi=M_X/(g_XQ_\Phi)$.
The vacuum stability constraint is denoted by red line; i.e., 
the region above the red line is allowed for $\alpha>0$,
and it is the other way around for $\alpha<0$. 
The perturbativity requirement is represented by blue line; i.e.,
the region below the blue line is allowed for $\alpha>0$, and it is the other way around 
for $\alpha<0$.
For $\alpha<0$, the region above the dotted black line is excluded by Eq.~(\ref{UB}).
Putting all together, for $\alpha>0$ the region between the red and blue lines is 
allowed while 
for $\alpha<0$ the region between the dotted black and blue lines is allowed.
It should be noted that since the coefficient of $\lambda_{H\Phi}$ in $\beta_{\lambda_H}$ 
is doubled in comparison with the real singlet case, the improvement of the vacuum stability by  the increase of $\lambda_{H\Phi}$ or, equivalently $\alpha$, is more effective.
However, unlike the general Higgs potential involving explicit $U(1)_X$ breaking terms, 
the EW vacuum cannot be stable up to Planck scale if $\alpha$ is exactly zero.

\begin{figure}[t]
\centering
\includegraphics[width=7.5cm]{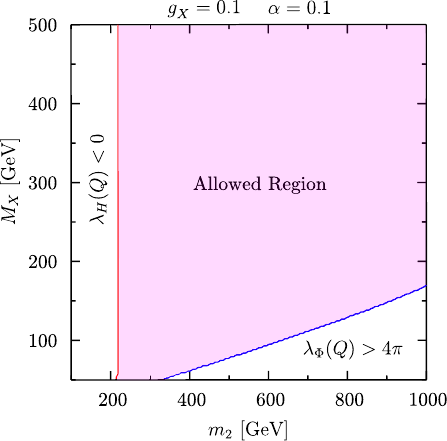}
\includegraphics[width=7.5cm]{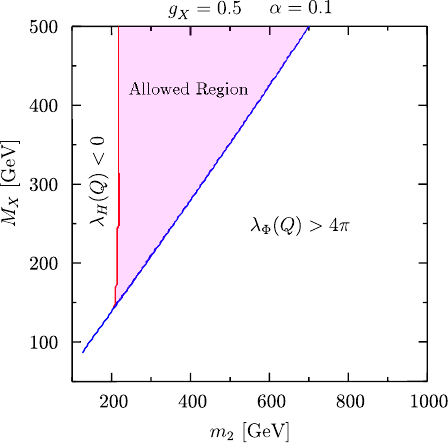}
\caption{
The vacuum stability and perturbativity constraints in the $M_X$-$m_2$ plane.
We set $g_X^{}=0.1$ (Left Panel) and 0.5 (Right Panel) with being $\alpha=0.1$.
}
\label{fig:RG_mX_m2}
\end{figure}
In Fig.~\ref{fig:RG_mX_m2}, we show 
the vacuum stability and perturbativity constraints in the $M_X$-$m_2$ plane.
We fix $\alpha=0.1$ varying $g_X^{}$, i.e., $g_X^{}=0.1$ (Left Panel) and 0.5 (Right Panel).
Once $g_X^{}$ is fixed, the small $M_X$ is realized by a small $v_\Phi$.
In such a case, the large $m_2$ is possible only by a large $\lambda_\Phi$ 
since $m_2\simeq \sqrt{2\lambda_\Phi} v_\Phi$ for a small $\alpha$. 
This explains the regions excluded by $\lambda_{\Phi}(Q)>4\pi$ in both plots.
Indeed, the $g_X^{}=0.5$ case yields the severer constraints. 
As for the vacuum stability constraint, the change of $g_X^{}$ has little effect on it,
which can be understood from the expression of $\beta_{\lambda_H}$, Eq.~(\ref{beta_lamH}).

\section{Conclusions}

In this paper, we revisited the Higgs portal vector dark matter including the hidden sector
Higgs field $\Phi$ that provides the vector dark matter mass.  Including the hidden 
sector Higgs field makes the model renormalizable and unitary. 
The constraint from direct detection cross section (XENON100) still allows a large parameter space in this model. 
On the contrary to some claims that the Higgs portal dark matter model
is strongly constrained by XENON100 data, we showed that the model is still viable. 
It is crucial to work with a model that is renormalizable, and not with effective 
lagrangian, as in the Higgs portal fermion DM model in Ref.~\cite{SFDM1,SFDM2}
Including the hidden sector Higgs field also improves the vacuum stability of the model
for $m_H = 125$ GeV upto the Planck scale as in Ref.~\cite{SFDM2}.  
Our model can be tested at colliders by searching for the 2nd Higgs boson and/or 
the signal strength of the 125 GeV Higgs boson.  It would take long in order to observe
the 2nd Higgs boson since its signal strength is smaller than 0.3. 
In our model, $r_i$ is universally suppressed relative to the SM case for all channels.  
This could be a useful criterion when the signal strengths  of 125 GeV Higgs boson 
are measured with smaller uncertainties. 
If $r_i$ is not universally suppressed or larger than one, then our model shall be excluded.

\acknowledgments

We are grateful to Yasaman Farzan for bringing Ref.~\cite{Farzan:2012hh}
to our attention.
This work is supported in part by NRF Research Grant 
2012R1A2A1A01006053 (PK and SB), and by SRC program of NRF 
Grant No. 20120001176 funded by MEST through 
Korea Neutrino Research Center at Seoul National University (PK).
WIP is supported in part by Basic Science Research Program through the National 
Research Foundation of Korea(NRF) funded by the Ministry of Education, 
Science and Technology(2012-0003102).


\appendix
\section{One-loop $\beta$ functions of Higgs quartic couplings}\label{sec:beta}
The renormalization group equation and the $\beta$ functions are given by
\begin{align}
\frac{d\lambda(t)}{d\log(Q)} = \beta_\lambda,
\end{align}
where
\begin{align}
\beta_{\lambda_H} &= \frac{1}{16\pi^2}
\left[
	24\lambda_H^2+\lambda_{H\Phi}^2-6 y_t^4
	+\frac{3}{8}\Big\{2g_2^4+(g_2^2+g_1^2)^2\Big\}
	-\lambda_H\Big\{3(3g_2^2+g_1^2)-12 y_t^2\Big\}
\right], \label{beta_lamH}\\ 
\beta_{\lambda_{H\Phi}} &= \frac{1}{16\pi^2}
\left[
	2\lambda_{H\Phi}(6\lambda_H+4\lambda_\Phi+2\lambda_{H\Phi})
	-\lambda_{H\Phi}\left\{\frac{3}{2}(3g_2^2+g_1^2)-6 y_t^2+6g_X^2Q_\Phi^2 \right\}
\right], \\
\beta_{\lambda_\Phi} &= \frac{1}{16\pi^2}
\Big[
	2(\lambda_{H\Phi}^2+10\lambda_\Phi^2+3g_X^4Q_\Phi^4)-12\lambda_\Phi g_X^2Q_\Phi^2
\Big], \\
%
\beta_{g_X} &= \frac{1}{16\pi^2}
	\frac{1}{3}g_X^3Q_\Phi^2.
\end{align}

\end{document}